# Genetic Variability of Splicing Sites

Dmitri V. Parkhomchuk

Email address: parkhomc@molgen.mpg.de , pdmitri@hotmail.com

**Abstract**

Splicing sites provide unique statistics in human genome due to their large number and reasonably complete annotation. Analyses of the cumulative SNPs distribution in splicing sites reveal a few interesting observations.

While a degree of the nucleotide conservation reflects on the SNPs density monotonically, no detectable changes in the SNPs frequencies spectrum were found. Semi-conserved nucleotide sites harbor transition mutations predominantly. We propose that such transition preference is caused by co-evolution of a site with corresponding binding agents. Since transitions in humans and similarly in other organisms are almost twice as frequent as transversions, this adaptation significantly lowers the mutation load.

**Results**

The sequences for approximately 330,000 splicing sites, which are annotated in NCBI build 35 human genome were extracted, along with all variable SNPs at these sites available at HapMap SNPs database for CEU population [1]. These HapMap SNPs were genotyped reasonably homogeneously at splicing sites. It was observed the excessive density of genotyping at some highly conserved nucleotides (GT and AG splicing sites) and exons, probably reflecting the hunt for functional variants. However this bias does not influence the main observations. In line with [2] the nucleotide sites variability and functional load were defined as illustrated on Fig. 1. The corresponding sequence logos are shown on Fig. 2. Nucleotide sites have broad distribution of functional load and, as it could be expected, the number of SNPs per site is proportional to the site variability. Exons have apparently lower SNPs density than introns, and for donor exon one can observe the traces of the increase of $3^{rd}$ codon position SNPs number because the large part of exons is in phase 0, i.e. $3^{rd}$ position is the last coding nucleotide before donor site. However no dependence of SNPs frequency was detected – the frequency distributions for neutral sites SNPs and SNPs at conserved sites are indistinguishable. It could be expected that conserved sites have more rare SNPs because purifying selection prohibits deleterious SNPs to rise in frequency. Although the statistics is rather large – hundreds of SNPs per nucleotide site, it was not possible to observe any differences. However HapMap sample size (60 unrelated individuals and 30 their children) may be insufficient to detect differences for rare SNPs.

Inspecting consensuses (Fig. 2) it is evident that the majority of semi-conserved sites have the next best-fit base as a transition mutation from the top base. The probability for a random pair of bases to be related by transition is 1/3 (Fig. 4), thus presumably this far from random pattern reflects optimisation for mutation load. Transitions are nearly twice as frequent as transversions in humans, thus when two best nucleotides for a given site are related by transition, a random mutation is more likely to be "synonymous" - not detrimental for site functioning. Fig. 4 demonstrates the apparent dependence of transversions to transitions ratio versus variability for the acceptor tail. This mechanism may work only for semi-conserved nucleotide site with functional load < 1 bit. At higher loads two equally good bases are impossible for an obvious

reason - 2 equally probable states give the entropy of 1 bit, thus for highly conserved sites there is indeed no significant preference for transitions (data not shown).

**Conclusion**

It is likely that most of non-coding functionality is not yet characterized and the amount of it in large genomes may be larger than coding part [4]. Arguably, the deciphering of non-coding functionality is the next large-scale hardest problem in genomics.

It seems that due to the generality of information theory, described observations could be usefully extended on non-coding sequences *en mass* [4].

Splicing sites, due to large statistics, may serve as a calibration reference for relative SNPs density versus functional load. For example, with functional load of 1 bit the SNPs density falls slightly more than twice, in comparison with neutral sequence. (Fig. 3) Apparently analogous decrease happens in orthologous sequences evolution. Adaptation of semi-functional sites (or better to say of their binding agents) for the prevalent transition mutations is analogous to the replacement-to-synonymous mutations (R/S) metric and can be equally useful in evolutionary analyses of non-coding sequences. This kind of adaptation to mutational bias appears to be quite ubiquitous as it affects the genetic code itself [3], where it is quite transparent for the $3^{rd}$ codon positions – nearly all transitions are strictly synonymous in contrary to transversions.

```
CTCGAGGGGTCCCGAGAGCACGCCTTCCAGATCACAGGTGTTTGGGATGCTTCC      GTGGGCAAGGCCCTCACCACGGTCCTGTCCTTGCAGATCCAGCACACGGAAGAC
CGGGGGGCCGCGGCGCTGCCACTCGGCACCCCACAGGTCAGTGCCGGGGACCC      GAGCCTCTGACGTCTGTGTCACCTTGTCCCAATCAGATCTGAATTCCAGAGAAC
TGCCTCGAGGGTCAAGCTGCAGCACCTGCCCGCACAGGTGGGTGGGAGGTGCGT      GCCTTTTCTCTGACGTTCCTTAACTCCCATGTGTAGCTATACACAGAATGGGAT
CATTTTCTCCGAGGAGCTGGACGGGCTTTGCTTCAAGGTGGGCCCCTCCCCACT      TGTGACTGTGATTCTTTCTCTCCTTTTGTCTTTCAGCAAACGGGGATTCAGCGT
GAGAAGGAGAAGCAGATCCGCTCCTTCCTGATTGAAGGTAGGGCCCTGACCCTG      CTAACTGTGCATCTTGGCATCTCCCTCGGCCACAGGGTTGGAAGCCCAGCGAG
GCCGCTGCCTGGTGCCGAGAGCTTCCCAGGGTCGCAGGTGAGGGGTCAATAGGC      GTCCCATGGTGGCCTGAGAATACCCTCTGCCCACAGGTCCCGAGCAGCCCGGCC
GCCATCCACCGTGGGCTGCTCCTCCCAGCACACACCGGTGAGCGCTTACGGGGT      GGCGGGAACGCCCCTCTGACCCCACCCCACCCCAGGCGTCCGGCACCCCAACA
CGGCCCTGTCACCCCACTGCACCTGGACCTGACCCAGGTGGGCCCAGCACACCC      TTCTTCTGAGAGCTTTATTTGTGAACCCTCTTGCAGTGTCACACTGAGTCCCCG
CCACCGCAGGGTCACAGATGTCCGGGGCCTGGAGGAGGTCAGGCCCCTGCTGGG      TCAGCCTTAGCCTGCTGGGGGGGCCTCTTTCCCAGGAGGGGAAGGGAAACCGG
GGATGGGGGCCGCAGCCCCAGACGCCCCTCAGCTTGTGAGTAGCAGCCCCCA       GCCGCTGCCTGAGGCCTGGTCTGCCACCCTCCGCAGGCAAGCCGGCGGAGCTGC
AGGGAAGGTTCGCCCGAACCCTGGCTGCCTCTGACAGGTGAGTAAGGATCCTGC      AGCCCGACCCAGCCACAGCTCCATGACCCGCCACAGTTTATCGGACAGACGGGC
TCGTGCCGTGGCCCACAGCCCACAGCCCACAGCCATGGTAAGGCAGATGTCACA      GCTGGGGGCCCCTCACGCCCTCTTTGTCGCTCAGCACCACTCCTTCTGGGTG
GCGGGGAGGCGGCACAGAGCCTGGCGGGCGGAGCTAAGGGCCTCCACCAGCATC      CCTGCAGCCTGCTGTGACCCCTCCCCTCCCGCAGGCCCTGGGCCGCGTCGGG
AGTGGTGCTGGGTCAGGCGGACCCAAGAGTCAGCCCGGTTAGGACCCATTTTAT     GCCCCTGGCCACCACTAACCTCAGCCCTGCCCCAGGCTCACTGAGCGTGGCCC
CTCCCCCACAGCTCCGGCCCTGCAGCCTGGCGAAGCCGTAAGTAACCCACATCA     GAGGGCCAGGGACTCACCTGCTGGCACTCTTGGCAGGTGGACACCAAGAACCAA
AGTAAACTCCCCAGGGCCTGCTACAGAAGAAACAAAGGTATCTCCCGGACCAGC     CCGGCGCCCTCCCTCTCCCACTTCCTCTCCTGTCAGGAACCAGCCCGAGGCCAC
AACAGAAGCTCAGGGACCCCGGGCGGCCACATCCTGGTAAGTCTCAGCAGGCC      GTCACCCCAGGTGACCACTGACTCCGCCCAGCAGGACGCCACTCGGACACG
CCTTCTTCTACATTGGAGGCAGCAACGGGGCCACAATGTGAGTAGCGGCCCTGG     AGTCCCAGGTCCCAGACCAACCTCCCTGCTCCACAGAGCTGTGAGAAAGATTCT
GTCAAGAGCCGAGACAGCTACGGCAGCTTCCGGGAAGGTAGCGGGAGCCGGCAC     GGTCCCTGGGCTGAGCCCGTCCCCACCCCTCCCAGGGCCGCTGTGACGTGAAC
GCAAGGCCAGCAAGGTGCCGGGGGGGTCCAGGCCAGGTGAGTCTGCCCCTGCC      GCGCCCGCCCTCACCGGCGTCTGTCCTGCCGCCAGCTACAGGCCTCGGGCCTC
.......................................................  .......................................................
```

donor         acceptor    ~330,000 annotated sites in human genome

$$\{f_A, f_C, f_G, f_T\} \quad \text{Base frequencies for a given nucleotide site}$$

$$H = - \sum_{i=A,C,G,T} f_i \log_2(f_i)$$

Site variability: $H$ (0-2 bits)
Site conservation (or functional load): $2-H$

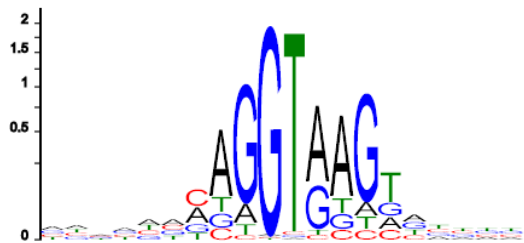

"Sequence logo" – total column height is proportional to conservation, letter heights to frequencies.

Fig. 1. Schema of data processing.

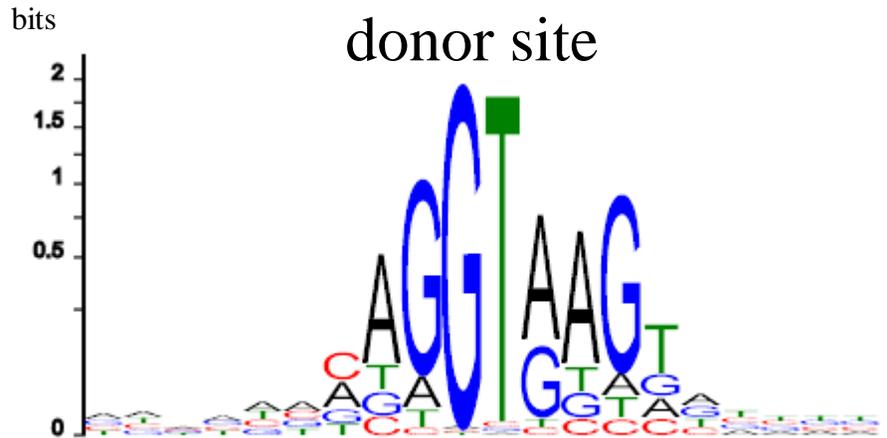

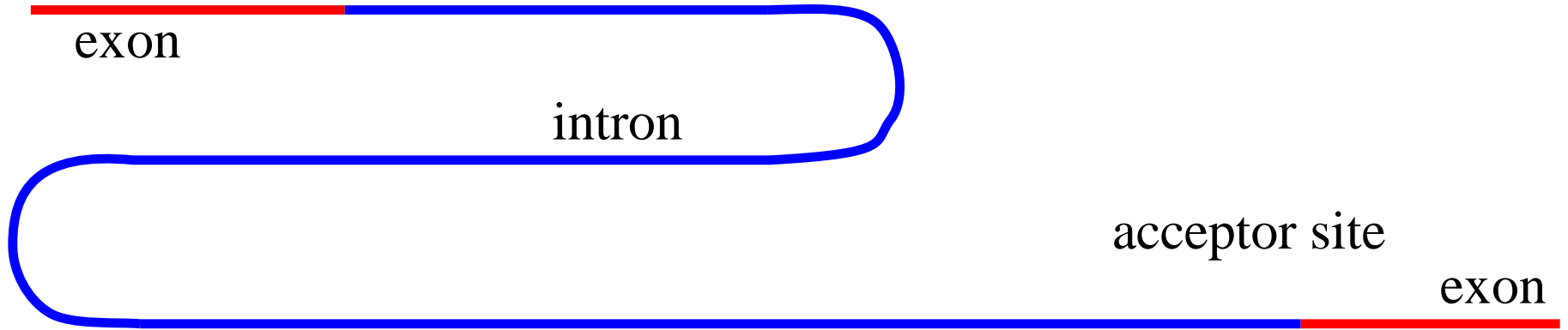

Fig. 2. Sequence logos for donor and acceptor sites. Note the non-random pattern of minor frequencies at the acceptor „tail".

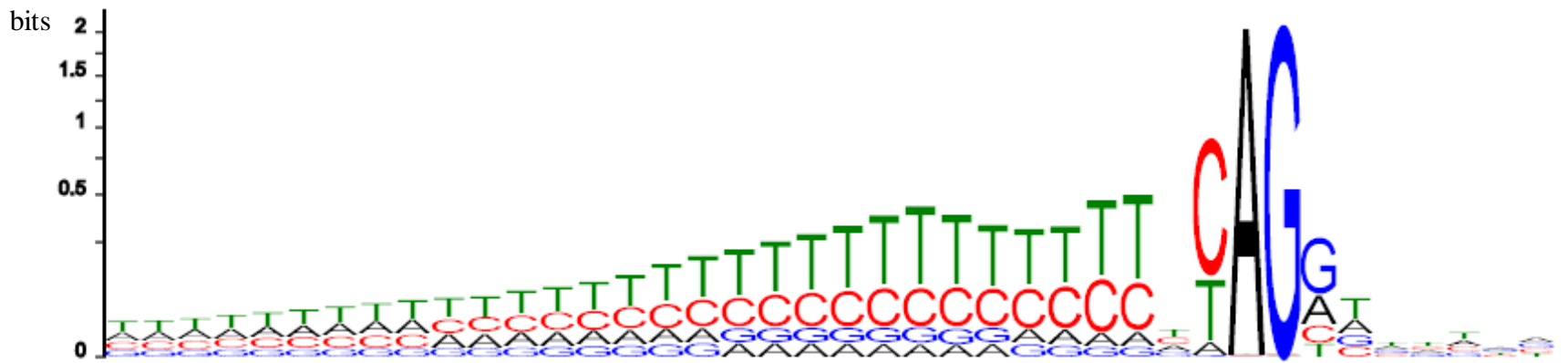

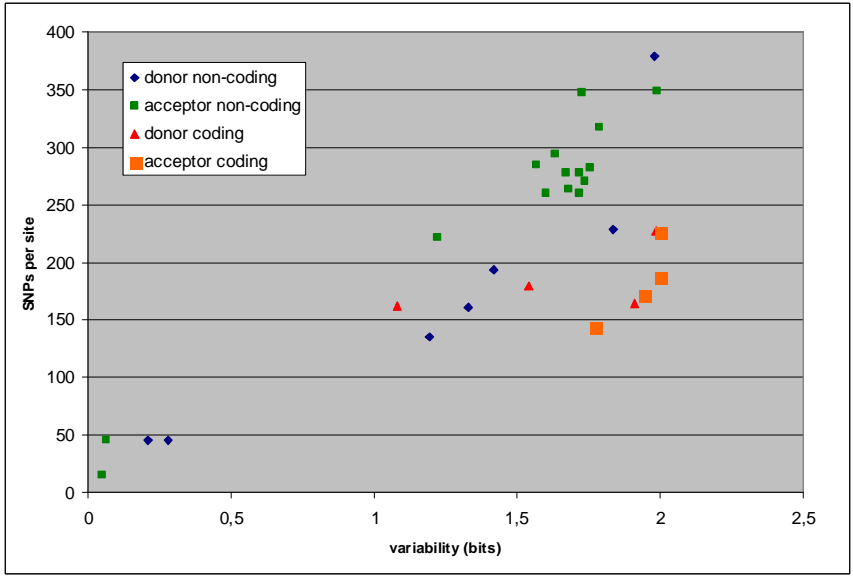

Fig. 3. SNPs per site vs. site variability. Exons have about 70% less variable SNPs than introns.

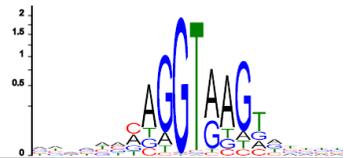
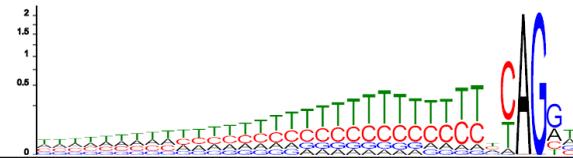
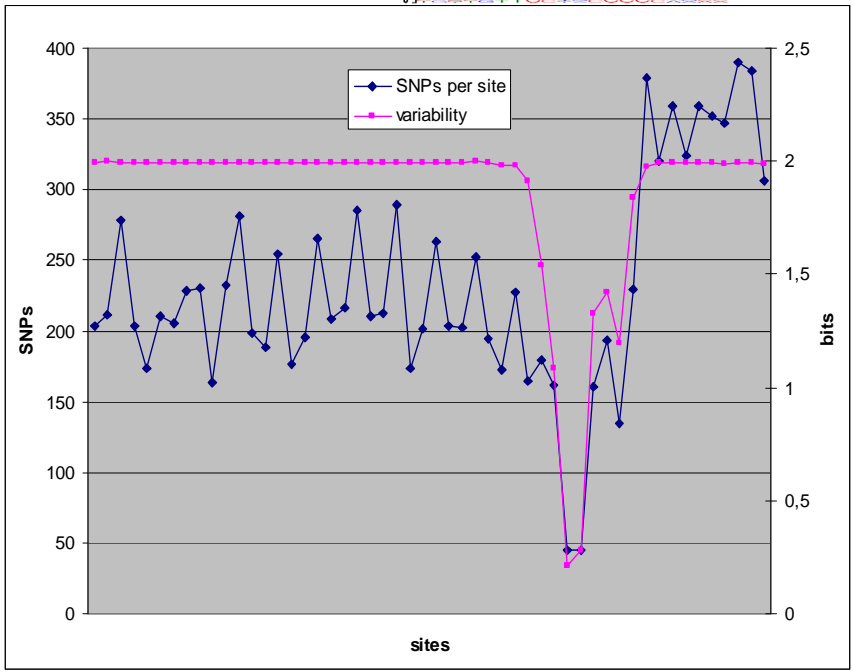
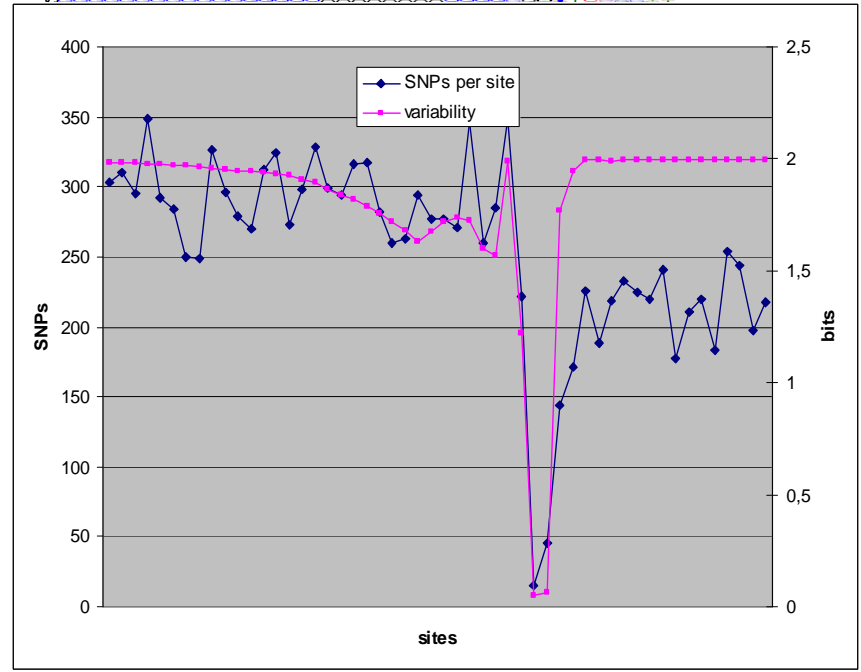

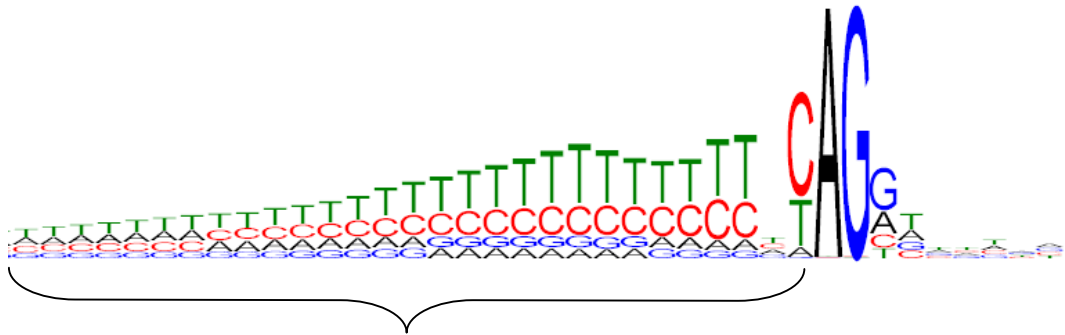
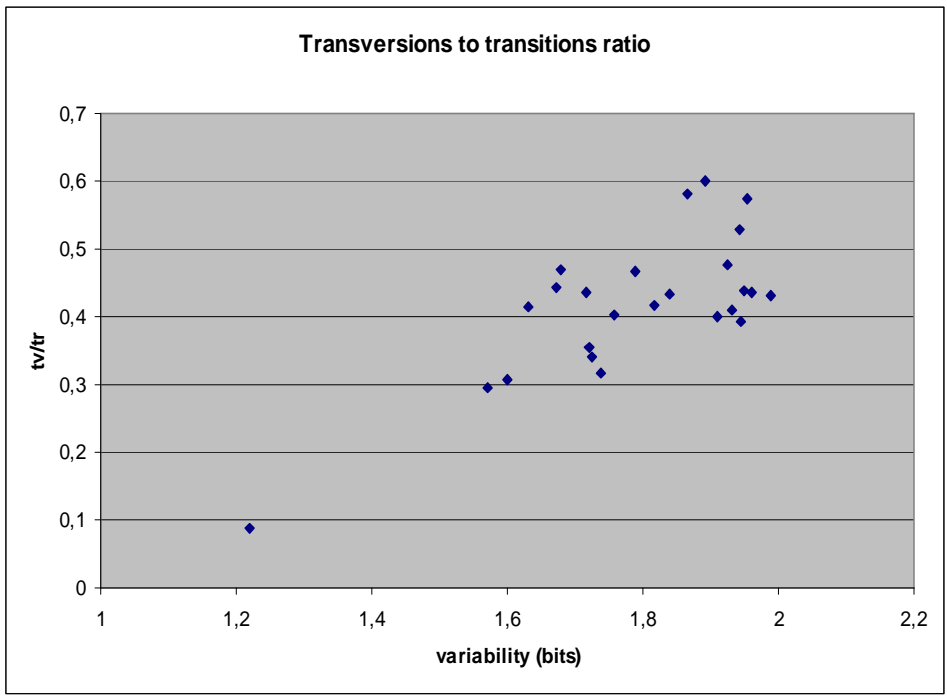

Fig. 4. Transversions to transitions ratio of SNPs at acceptor site tail vs. nucleotides variability.

| | | mutation rates |
|---|---|---|
| A-G  C-T | transitions | 2N |
| A-C  C-G  ... | transversions | N |
| A-T  C-A  ... | | |